\documentclass{appolbaXv}
\usepackage{graphicx}
\usepackage{amssymb}
\newcommand{\eq}{\begin{equation}}
\newcommand{\eqx}{\end{equation}}
\newcommand{\eqn}{\begin{eqnarray}}
\newcommand{\eqnx}{\end{eqnarray}}
\newcommand{\bi}{\begin{itemize}}
\newcommand{\ei}{\end{itemize}}
\newcommand{\nn}{\nonumber}
\newcommand{\ra}{\rangle}
\newcommand{\la}{\langle}
\newcommand{\bz}{\bar{z}}
\newcommand{\si}{\sigma_I}
\newcommand{\sr}{\sigma_R}

\begin{document}
\title{\bf Beyond complex Langevin equations I: two simple examples%\thanks{Presented at the Light Cone Cracow 2012, 8-13 July 2012, Krakow, Poland}%
}
\author{Jacek Wosiek\thanks{e-mail: wosiek@th.if.uj.edu.pl}
\address{M. Smoluchowski Institute of Physics, Jagellonian University\\
\L ojasiewicza St. 11, 30-348 Krakow}
%the Name(s) of other Author(s)
%\address{affiliation}
}
\maketitle
\begin{abstract}
By introducing a second complex variable, the integral relation between a complex density and the corresponding positive distribution is derived. Together with the positivity and normalizability conditions, this sum rule allows to construct explicitly  equivalent pairs of distributions in simple cases discussed here. In particular the well known solution for a complex gaussian distribution is generalized to an arbitrary complex inverse dispersion parameter. This opens a possibility of positive representation of Feynman path integrals directly in the Minkowski time.
\end{abstract}
%\PACS{PACS numbers come here}
  
\section{Basics}
Generally quantum averages result from weighting observables with complex amplitudes rather than with positive
probabilities. The technique colloquially referred to as Complex Langevin can in principle be used to replace this by a standard, statistical averaging over a suitably defined stochastic process. The method was proposed long time ago \cite{P,Kl}, but recently has attracted a new wave of interest, especially in studies of quantum chromodynamics at finite chemical potential \cite{S1,S2}. Still, contrary to the real Langevin approach, there is no general proof of the convergence \cite{S3,S4} and the evidence for the success is somewhat limited \cite{Ph,Bl}. 

In this article  the positive probabilities, which replace complex weights, will be directly constructed  (i.e. without any reference to  stochastic processes and/or Fokker-Planck equations). Thus above mentioned  difficulties are avoided albeit in a few simple test cases. 
Building on this the positive representation for Feynman path integrals could be derived. This is done in Ref.~\cite{Ja} for some typical 
quantum mechanical applications.
  
The essence of the complex Langevin approach is summarized by the following relation
\eqn
\frac{\int f(x) e^{-S(x)} dx}{\int e^{-S(x)} dx } = \frac{\int \int f(x+i y) P(x,y) dx dy }{\int \int P(x,y) dx dy}. 
\label{B1}
\eqnx
For complex action, $S(x)$, the left hand side (LHS) does not have statistical meaning, however the right hand side (RHS) does, since $P(x,y)$ is the
well defined distribution of the long stochastic process associated with $S(x)$.  Precise form of the Langevin equation is not relevant here. Suffices to say that for real action the Langevin process is real -- the density $P(x)$ is concentrated on the real axis. It satisfies the Fokker-Planck (FP) equation whose solution
converges to $e^{-S(x)}$ for large Langevin time. On the other hand,  for complex actions the stochastic trajectory is driven into the complex plane. The density $P(x,y)$ satisfies the  FP equation in two variables, however the asymptotic (in the Langevin time) behaviours of solutions are not known in general and their relations to the original complex actions are not clear \cite{AY,AFP,HW}. 

Nevertheless the complex Langevin approach  {\em is known} to work  as has been proven in the gaussian case:
\eqn
S_g(x)=\frac{1}{2}\sigma x^2,\;\;\;\;\sigma=\sr+i \si, \;\;\;\sr > 0.\label{S}
\eqnx
The large time asymptotics of the solution of the corresponding FP equation has been given in \cite{AY} 
\eqn
P_g(x,y)=\exp{\left(-\sr (x^2+2 r x y + (1+2 r^2) y^2)\right)},\;\;\;\; r=\frac{\sr}{\si}, \label{P}
\eqnx
\eqn
\int_{R^2} P_g(x,y)=\frac{\pi}{\sr\sqrt{1+r^2}} , \nn  %\label{GN}
\eqnx
and thoroughly analysed in the literature \cite{DH,HP}.

To see the validity of (\ref{B1}), e.g. for polynomial observables, consider the generating function
\eqn
G_{LHS}(t)= \frac{\int_{-\infty}^{\infty} e^{t x} e^{-S_g(x)} dx}{\int_{-\infty}^{\infty} e^{-S_g(x)} dx } =\exp{\left(\frac{t^2}{2\sigma}\right)}, \label{B2}
\eqnx
and the average from the RHS of (\ref{B1}) 
\eqn
G_{RHS}(t)=\frac{\int_{-\infty}^{\infty}\int_{-\infty}^{\infty}  dx dy e^{ t (x + i y) } P_g(x,y)}{\int_{-\infty}^{\infty}\int_{-\infty}^{\infty}  dx dy  P_g(x,y)}, \nn  \label{B5}
\eqnx
which indeed agrees with (\ref{B2}).

Summarising: the complex Langevin approach can in principle be used to perform simulations with ``complex distributions". 
However, in practice, extending 
the stochastic process into a complex plane encounters difficulties. Asymptotic solutions of the two dimensional Fokker-Planck equation
are generally not known and cannot be simply constructed from the complex action. Moreover, the random walk wanders  often far  
into the imaginary direction and may run away or converge to the wrong answer.

\section{Generalization}
On the other hand we do not really need to generate the positive two dimensional distribution with the stochastic process in the complex plane.
The only and the real problem is to find a positive distribution which satisfies (\ref{B1}). Given $P(x,y)$ one can generate it with other
methods. 

Therefore, we propose to avoid difficulties of the complex random walk and concentrate instead on constructing $P(x,y)$ directly, using
eq.(\ref{B1}) as a guide. To this end rewrite the RHS of (\ref{B1}) in terms of two, holomorphic and antiholomorphic, variables
\eqn
\;\;\;z=x+iy,\;\;\;\bz=x-iy,     \label{xy}
\eqnx
\eqn
 \frac{\int_R \int_R f(x+i y) P(x,y) dx dy }{\int_R \int_R P(x,y) dx dy}=\frac{\int_{\Gamma_z} \int_{\Gamma_{\bz}} f(z) P(z,\bz) dz d\bz }{\int_{\Gamma_z} \int_{\Gamma_{\bz}} P(z,\bz) dz d\bz}. \label{Pz}
\eqnx
Now, continue analytically the complex density on the LHS of  (\ref{B1}) from the real axis into the complex plane
\eqn
\rho(x)=e^{-S(x)} \longrightarrow \rho(z), \nn
\eqnx
rotate the contour of integration on the LHS of (\ref{B1}), $R\rightarrow \Gamma_z$, and then seek to satisfy the relation
\eqn
 \frac{\int_{\Gamma_z} f(z) \rho(z) dz}{\int_{\Gamma_z} \rho(z) dz} = \frac{\int_{\Gamma_z} \int_{\Gamma_{\bz}} f(z) P(z,\bz) dz d\bz }{\int_{\Gamma_z} \int_{\Gamma_{\bz}} P(z,\bz) dz d\bz}. \label{B3}
\eqnx
This will be the case provided
\eqn
\rho(z)=\int_{\Gamma_{\bz}} P(z,\bar{z}) d \bar{z}.  \label{Pro}
\eqnx
That is, we will look for the distribution $P(z,\bz)$, which: (1) upon integration over $\bz$ reproduces the analytic continuation $\rho(z)$, 
and (2) is positive and normalizable when expressed in terms of real and imaginary parts $x$ and $y$.   Given that, we will have found
the positive representation for the LHS of (\ref{B3})
\eqn
 \frac{\int_{\Gamma_z} f(z) \rho(z) dz}{\int_{\Gamma_z} \rho(z) dz} = \frac{\int_{R^2} f(x+iy) P(x,y) dx dy }{\int_{R^2} P(x,y) dx dy}.\nn  % \label{B4}
\eqnx
The integral on the RHS is over the whole $(x,y)$ plane (at least in the cases considered here), while the contours $\Gamma_z$  and $\Gamma_{\bz}$  
have to be within domains determined by parameters of both distributions. For a range of parameters
a domain for $\Gamma_z$ contains the real axis and then Eq.(\ref{B1}) can be established. 

It is shown below that this program can in fact be carried through quantitatively, at least in few physically interesting cases, already providing 
some novel results.

\section{Generalized gaussian model}
A more general than (\ref{P}) positive distribution can be derived if we start from a generic quadratic action for (\ref{Pz}) in two complex variables $z$ and $\bz$
\eqn
S(z,\bar{z})&=& a^* z^2 + 2 b z \bz + a \bz^2, \nn  %\label{Sa}
\eqnx
with an arbitrary complex $a=\alpha+i\beta$ and real $b=b^*$.
In terms of real and imaginary parts (\ref{xy})
\eqn
S(x,y)= 2(b+\alpha) x^2 + 4\beta x y + 2(b-\alpha) y^2\label{Sb},
\eqnx
and gives the positive and normalizable (for real $x$ and $y$) distribution 
\eqn
P(x,y)= \exp{\left(-S(x,y)\right)}, \label{Pb}
\eqnx
provided $b > |a|$, since the two eigenvalues of (\ref{Sb}) 
\eqn
\lambda_{\pm}=2 (b\pm |a|). \nn
\eqnx
At the same time the normalization reads
\eqn
\int_{R^2} dx dy P(x,y) = \frac{\pi}{2\sqrt{b^2-|a|^2}}.\label{Pc}
\eqnx
On the other hand, integrating 
\eqn
P(z,\bz)=\frac{i}{2}P(x,y)=\frac{i}{2}\exp{\left(-S(z,\bz)\right)}, \nn
\eqnx
 as in (\ref{Pro}), gives
\eqn
\rho(z)=\int_{\Gamma_{\bz}} P(z,\bz) d \bz = 
%\frac{1}{2}\sqrt{\frac{\pi}{-a}}\exp{\left(\frac{b^2-|a|^2}{|a|^2}a^* z^2\right)} \nn \\ 
  \frac{1}{2}\sqrt{\frac{\pi}{-a}}\exp{\left( - s z^2\right)},\;\;\;s=\frac{|a|^2-b^2}{a}. \label{ip2}
\eqnx
which is properly normalized in view of (\ref{Pc}).
The contour $\Gamma_{\bz}$ depends on a phase of a complex parameter $a$ and is chosen such that the integral converges. This choice also determines unambiguously the phase of $-a$.

With $a$ and $b$ parametrized by 
\eqn
b=\frac{\sigma_R}{2}(1+r^2) , \;\;\;\;  \alpha=-\frac{\sigma_R}{2} r^2 , \;\;\;\; \beta=\frac{\sigma_R}{2} r,  \;\;\;\;\sigma_R>0, \nn % \label{con}
\eqnx
equations (\ref{Sb}) and (\ref{ip2}) reproduce the original gaussian model, i.e. (\ref{P}) and (\ref{S}) respectively. However (\ref{Sb}) gives a more general, positive and normalizable  probability. In fact the generalized model (\ref{Sb}) realizes the positive representation of the gaussian (\ref{ip2}) for any complex value of the slope, $s$, or equivalently $a$, $b > |a|$.

A complex gaussian, e.g. $e^{- s z^2}, s,z \in C$, is integrable only along a family of contours contained in a wedge specified by a phase 
of $s$. However its moments can be analytically continued to any complex $s$. The point of (\ref{Sb},\ref{Pb}) is that it provides a positive and normalizable integral representation for this continuations at arbitrary complex $s$.
In another words: even though the complex density $\rho$ was derived and is integrable only along particular family of contours for a given $a$, 
the positive density $P(x,y)$ exists and is integrable for all $a\in C$.   

It is a simple matter to check the equivalence of (\ref{Pb},\ref{Sb}) and (\ref{ip2}), e.g. by calculating generating function (\ref{B5}) with both representations.
Here we illustrate this only for the second moment. In the matrix notation the action (\ref{Sb}) reads
\eqn
S(x,y)=X^T M X,\;\;\;X^T=(x,y). \nn
\eqnx
Therefore
\eqn
\la (x+i y)^2 \ra_P=\frac{1}{2}\left(M^{-1}_{11}-M^{-1}_{22} + 2 i M^{-1}_{12}\right)=
-\frac{1}{2}\frac{\alpha+i\beta}{b^2-|a|^2}, \nn
\eqnx
which indeed is identical to the average over the complex density (\ref{ip2})
\eqn
\la z^2 \ra_{\rho} = \frac{1}{2 s}. \nn
\eqnx

To conclude this Section we discuss two interesting special cases.

 For real and negative $s$, the complex density blows up along the real axis. On the other hand the distribution $P(x,y)$ is positive and normalizable at $\alpha>0$ and  $\beta=0$ producing the correct average over the "divergent" distribution $\rho$. This explains a ``striking example" observed in the literature \cite{AS}, namely that, upon change of variables, the complex Langevin simulation based on (\ref{P})  actually has the correct fixed point also for negative ${\cal R}e\;\sigma$. The answer is that the positive distribution (\ref{P}) used until now is part of a richer structure (\ref{Sb}), which accommodates negative $\sigma_R$ as well.
 
 Similarly, the complex density $\rho(z)$ for purely imaginary $s$ is readily represented by the positive distribution $P(x,y)$, which is perfectly well defined at $\alpha=0$ and arbitrary $\beta$, as long as $|\beta|<b$. This opens an exciting possibility of positive representations for Feynman path integrals directly in the Minkowski time. Such a construction is presented in detail in \cite{Ja}.
 
 In both cases the original density (\ref{P}) does not exist.

\section{A quartic model}

Another possible solution obtains if we start from the action
\eqn
S_4(z,\bz)=(a^* z^2+2 b z \bz + a \bz^2)(c^* z^2+2 d z \bz + c \bz^2), \nn  %\label{S4}
\eqnx
with complex $a$ and $c$ and real $b \gtrless |a|$ and $d \gtrless |c|$. The density $P(x,y)$ is again positive and normalizable on the $x,y$ plane.
To derive $\rho(z)$ introduce an arbitrary shift parameter $e$ and change the variables. This gives
 \eqn
S_4(z,\bz=u-e z)=A_0 z^4 + A_1 z^3 u + A_2 z^2 u^2 + A_3 z u^3 + A_4 u^4,\nn
\eqnx
with
\eqn
A_4&=& a c,\nn\\
A_3&=&2(a d + b c)- 4 a c e,\nn\\
A_2&=&a^* c + c^* a + 4 b d - 6 e ( a d + b c) + 6 a c e ^2,\nn\\
A_1&=&2(b c^* + a^* d) - 2 e (a^* c+ a c^*-4 b d) + 6 e^2 (b c + a d)- 4 e^3 a c,\nn\\
A_0&=&(a^* - 2 b e + a e^2 )(c^* - 2 d e + c e^2).\nn
\eqnx
Now choose $e$ such that $A_3=0$. The coefficients become
\eqn
A_4&=& a c,\nn\\
A_2&=&\frac{1}{2 a c}\left(2 a^2(|c|^2-d^2)+2 c^2(|a|^2 - b^2)-(a d - b c)^2\right),\nn\\
A_1&=&\frac{1}{a^2 c^2}\left( a d - b c \right)\left(a^2(|c|^2-d^2)-c^2(|a|^2-b^2) \right),\nn \\
A_0&=&\frac{1}{16 a^3 c^3}\left(4 c^2(|a|^2-b^2)+(a d - b c)^2\right)\left(4 a^2(|c|^2-d^2)+(a d - b c)^2\right).\nn
\eqnx
Then $A_1$ can be also eliminated setting
\eqn
c=\frac{d}{b} a, \nn
\eqnx
which essentially reduces $S_4(z,\bz)$ to a square. Remaining coefficients simplify
\eqn
A_4&=&\frac{d}{b}a^2,\nn \\
A_2&=&2 \frac{d}{b} \left(|a|^2- b^2\right),\nn \\
A_0&=&\frac{d}{b} \left(|a|^2- b^2\right)^2 \frac{1}{a^2}.\nn
\eqnx
The complex density $\rho_4(z)$ can be then obtained in a closed form as
\eqn
\rho_4(z)&=&\frac{i}{2}\int_{\Gamma_{\bz}} d \bz  e^{- S_4(z,\bz)}\nn\\&=&\frac{i}{2}\exp{\left(- A_0 z^4\right) } \int_{\Gamma_u} d u \exp{\left(- A_4 u^4 - A_2 z^2 u^2\right)} \label{prop}\\
&=&\frac{i}{2}\left(\frac{b}{2 d a^2}\right)^\frac{1}{4}  \exp{\left(-\sigma z^4\right) } \left(\sigma z^4 \right)^\frac{1}{4} 
K_{\frac{1}{4}}\left(\sigma z^4\right),\nn
\eqnx
with an arbitrary complex
\eqn
\sigma=\frac{d (b^2-|a|^2)^2}{2 b a^2}.\nn % \label{sQ}.
\eqnx
All contours (here and below) are such that the integrals exists. Basically one can choose straight lines with slopes determined by 
the phase of $a$. 

It is a simple exercise to show that normalization of both densities is the same:
\eqn
\int_{\Gamma_z} \rho_4(z) dz= \frac{i}{2}\left(\frac{b}{2 d a^2}\right)^\frac{1}{4}\int_{\Gamma_z}   \exp{\left(-\sigma z^4\right) } \left(\sigma z^4 \right)^\frac{1}{4} K_{\frac{1}{4}}\left(\sigma z^4\right)=\nn\\
\frac{\pi^{\frac{3}{2}}}{4}\sqrt{\frac{b}{d(b^2-|a|^2)}}=\nn\\
\int_{R^2} dx dy e^{- 4 \frac{d}{b} \left( (b+\alpha)x^2 +2 \beta x y +(b - \alpha)y^2 \right)^2 }=\int_{R^2} dx dy P_4(x,y).\nn
\eqnx
The difference however, being that while on the LHS the density $\rho$ is in general complex and contour $\Gamma_z$ has to be adjusted depending on a phase of $\sigma$, the integral on the RHS is always over $R^2$ and the density $P_4(x,y)$ is positive and normalizable for all complex $\sigma$. The same applies for higher moments:
\eqn
\int_{\Gamma_z} z^n \rho_4(z) dz = \int_{R^2} dx dy (x+i y)^n P_4(x,y).  \nn
\eqnx
In fact the construction works for a larger range of parameters that in the gaussian case since the condition
$|a| < b$ can be released.

The density (\ref{prop}) has the simple leading asymptotics 
\eqn
  \rho_4(z)\sim e^{\left(- 2 \sigma z^4 \right)} ,\;\;\; z \longrightarrow \infty, \nn
\eqnx
and therefore might be of some practical interest (e.g. in optimizing some reweighting algorithms). The main point of this example is however, that the original idea, namely constructing positive representations by introducing a second variable, seems to be general and points towards existence of some unexplored yet structures.

Obviously there is a lot of freedom in choosing an initial action. It remains to be seen to what extent this freedom allows to derive  complex densities of wider physical interest. 

\section{Summary and outlook}
Simulating complex distributions with the complex Langevin technique still rises some theoretical doubts and practical difficulties. 
These problems can be avoided by direct construction of corresponding positive densities. It turns out that introducing 
a second complex variable allows to carry this programme in practice at least in the two cases discussed in this article. 

One is the gaussian distribution with a complex inverse dispersion parameter $\sigma$. The corresponding positive density, in the case of positive $ {\cal R}e\; \sigma$, is known for a long time. Within the present approach  this classical result is generalized to arbitrary $\sigma \in C$. In particular the equivalent positive
and normalizable density exists for purely imaginary $\sigma$. Thereby an intriguing possibility of positive representations for Feynman path integrals, directly in the Minkowski time, emerges \cite{Ja}.  

The second example deals with
the specific quartic action. Again the equivalent, positive density is explicitly constructed and is valid even in the larger range of parameters, than for the gaussian case. 

Of course plenty of questions require further studies. First, we would like to invert the procedure followed here, namely instead of
deriving the complex density $\rho$ by integrating a positive ansatz for $P$, one really needs to derive the latter from the former. At first sight this may not have a unique answer. While this is not the main problem (any answer would do), the opposite question poses the real challenge. Namely whether, and under which conditions, $P$ exists at all. Also, it remains to be seen whether and how the positivity, normalizability and the integral sum rule (\ref{Pro}) allow to determine $P$ in practice for more general cases.

Keeping these reservations in mind, a host of further extensions and applications suggest itself. To name a few: generalization for other nonlinear actions, compact integrals, many degrees of freedom, path integrals, field theory etc. We are looking forward for further studies of these issues.

\vspace*{.5cm}

\noindent{\em Acknowledgements} \   Existence of positive representations for complex, gaussian and more general, densities has been discussed in \cite{Sa,We}. The projection relation employed in \cite{Sa} bears some similarity to the condition (\ref{Pro}).

The author would like to thank Owe Philipsen for the discussion.

\end{document}